\documentclass{article}
\usepackage[preprint,nonatbib]{neurips_2025}
\usepackage{graphicx} 
\usepackage{float}
\usepackage{amsmath}
\usepackage[backend=biber]{biblatex}
\usepackage{appendix}
\usepackage[colorlinks=true, linkcolor=blue, urlcolor=blue]{hyperref}
\usepackage[utf8]{inputenc} 
\usepackage[T1]{fontenc}    
\usepackage{hyperref}       
\usepackage{url}            
\usepackage{booktabs}       
\usepackage{amsfonts}       
\usepackage{nicefrac}       
\usepackage{microtype}      
\usepackage{xcolor}         
\title{Solvaformer: an SE(3)‑equivariant graph transformer for small molecule solubility prediction}
\workshoptitle{Non-Euclidean Foundation Models and Geometric Learning Workshop}
\author{%
  Jonathan Broadbent \\
  Sanofi, Digital R\&D\\
  Toronto, ON M5V 0E9 \\
  \And
  Michael Bailey \\
  Sanofi, Digital R\&D\\
  Toronto, ON M5V 0E9 \\
  \And
  Mingxuan Li\\
  Sanofi, Digital R\&D\\
  Cambridge, MA 02141\\
  \And
  Abhishek Paul\\
  Sanofi, CMC Synthetics\\
  Cambridge, MA 02141\\
  \And
  Louis De Lescure\\
  Sanofi, CMC Synthetics\\
  Cambridge, MA 02141\\
  \And
  Paul Chauvin\\
  Sanofi, Digital R\&D\\
  Barcelona, 08016, Spain\\
  \And
  Lorenzo Kogler-Anele\\
  Sanofi, Digital R\&D\\
  Toronto, ON M5V 0E9 \\
  \And
  Yasser Jangjou\\
  Sanofi, CMC Synthetics\\
  Cambridge, MA 02141\\
  \And
  Sven Jager\\
  Sanofi, Digital R\&D\\
  Frankfurt, 65929, Germany\\
  \texttt{sven.jager@sanofi.com}
}

\date{\today}
\DeclareUnicodeCharacter{202F}{\,}
\begin{document}

\maketitle
\begin{abstract}
Accurate prediction of small molecule solubility using material-sparing approaches is critical for accelerating synthesis and process optimization, yet experimental measurement is costly and many learning approaches either depend on quantum-derived descriptors or offer limited interpretability. We introduce \emph{Solvaformer}, a geometry-aware graph transformer that models solutions as multiple molecules with independent $SE(3)$ symmetries. The architecture combines \emph{intramolecular} $SE(3)$-equivariant attention with \emph{intermolecular} scalar attention, enabling cross-molecular communication without imposing spurious relative geometry. We train Solvaformer in a multi-task setting to predict both solubility ($\log S$) and solvation free energy ($\Delta G_{\mathrm{solv}}$), using an alternating-batch regimen that trains on quantum-mechanical data (CombiSolv-QM) and on experimental measurements (BigSolDB~2.0). Solvaformer attains the strongest overall performance among the learned models and approaches a DFT-assisted gradient-boosting baseline, while outperforming an EquiformerV2 ablation and sequence-based alternatives. In addition, token-level attention produces chemically coherent attributions: case studies recover known intra- vs.\ inter-molecular hydrogen-bonding patterns that govern solubility differences in positional isomers. Taken together, Solvaformer provides an accurate, scalable, and interpretable approach to solution-phase property prediction by uniting geometric inductive bias with a mixed dataset training strategy on complementary computational and experimental data.
\end{abstract}

\section{Introduction}
Small molecule synthesis is an essential operation in the development of therapeutics. The synthesis pathway involves a sequence of reactions, each yielding intermediate products. Rapid optimization of a synthesis process is critical to enable cost-effective and timely manufacturing of the final active pharmaceutical ingredient (API)~\cite{de_ruyter_optimization_chemrev}.

While many variables influence reaction speed and yield, here we focus on solubility. The choice of solvent impacts reaction kinetics, equilibria, and overall process efficiency; reactions often require solvents in which intermediates are highly soluble and sometimes anti-solvents to facilitate product isolation~\cite{byrne_tools_2016, zhang_prediction_solubility}.

A major limitation is that intermediate solutes are novel and lack prior characterization, and are available only in small amounts due to their transient role in the synthesis pathway. Consequently, many experimental solubility measurements are not feasible. This creates a strong need for predictive solubility models to estimate small-molecule solubility with minimal experimentation~\cite{zhang_prediction_solubility}.

Predicting small-molecule solubility is a major challenge in chemo-informatics and process chemistry. Despite decades of research and the publication of multiple large curated datasets, recent benchmarking reveals that even state-of-the-art models often fail to generalize reliably outside their training domains due to overfitting, inconsistent data quality, and limited applicability domains~\cite{llompart2024will}.

\subsection{DFT-based and COSMO-RS Models}

Density Functional Theory (DFT) is a method used to approximate quantum mechanical properties of molecular systems~\cite{parr1989density}. It forms the basis for physics-driven solubility models such as COSMO-RS (COnductor-like Screening MOdel for Real Solvents), which estimates solvation free energies using DFT-derived surface polarization charges~\cite{klamt2005cosmo}. These first-principles approaches provide physically meaningful predictions without relying on experimental data. 
While DFT methods such as COSMO-RS provide accurate estimates of solvation free energy, they do not directly predict solubility, which also depends on solid-state effects. Machine learning models trained on DFT-derived inputs can learn this mapping, thereby improving solubility prediction while retaining thermodynamic interpretability~\cite{klamt2002cosmo_rsol}. However, DFT based models first require the direct computation of molecular conformers which can take up to 10 hours for large molecules. Therefore, their high computational inference cost makes them unsuitable for large-scale screening~\cite{kastenholz2006cosmo}.

\subsection{Message-Passing Neural Networks (MPNNs)}

MPNNs represent an alternative data-driven method that operates directly on molecular graph structures. As first formalized by Gilmer et al.~\cite{gilmer2017mpnn}, these models iteratively exchange information between atoms (nodes) and bonds (edges) to build learned representations. MPNNs have shown strong performance on molecular property prediction tasks, including aqueous solubility, often outperforming classical descriptor-based and SMILES-string methods~\cite{panapitiya2022evaluation}. However, MPNNs lack explainability, which likely leads to caution of adoption in real-world scenarios.

Both DFT-derived and MPNN models exhibit trade-offs: DFT-based models have strong physical basis but poor scaling, while MPNNs offer scalability and performance but lack explicit physical reasoning. The need to integrate large-scale quantum datasets (e.g., CombiSolv-QM) and experimental data (e.g., BigSolDB 2.0) in a unified architecture without compromising interpretability or scaling capability motivates our choice of an attention‑based graph transformer architecture with multi-task outputs for both \(\Delta G_{\rm solv}\) and \(\log S\).

\subsection{SolvBERT}

SolvBERT is a transformer-based model that treats solute–solvent complexes as combined SMILES sequences, applying NLP-style encoding to molecular interactions~\cite{yu2023solvbert}. Unlike graph-based models that process solute and solvent separately, SolvBERT ingests the concatenated SMILES of the complex and converts them into contextualized embeddings using a BERT backbone pretrained in an unsupervised masked language modeling manner on a large computational dataset (CombiSolv‑QM)~\cite{yu2023solvbert}. With this setup, the self-attention network is able to learn interactions between solute and solvent. Following pretraining, the model is fine-tuned either on experimental solvation free energy or solubility datasets, demonstrating strong performance across both tasks.

Empirical evaluations show that SolvBERT achieves solvation free energy predictive accuracy comparable to state-of-the-art graph-based models like DMPNN. It also surpasses hybrid graph-transformer architectures such as GROVER when predicting solubility on out-of-sample solvent–solute combinations~\cite{yu2023solvbert}. The unsupervised pretraining enables better internal clustering of molecular systems (via TMAP visualization), supporting enhanced generalization despite varied fine-tuning targets.

\subsection{EquiformerV2}
Equiformer~\cite{thomas2022equiformer} is an SE(3)-equivariant transformer architecture designed for 3D molecular and material data. It extends existing transformer models by incorporating geometric attention mechanisms that respect the symmetries of three-dimensional space: namely rotation and translation. Equiformer is analogous to an ordinary graph transformer in the following sense:
\begin{itemize}
    \item Instead of weights and activations taking scalar values, they take values in an SO(3) representation space. These representations are equivalent to \emph{spherical harmonics} (also known as \emph{orbitals}), so a weight or activation can be seen as an approximated function on the sphere $S^2$.
    \begin{itemize}
        \item When a representation vector $f$ is decomposed into irreducible representations (i.e., different angular frequencies) $f_\ell$, its `rotations` correspond to Wigner D-matrices:
        $$
        f_{\ell} \mapsto D^\ell(R)\,f_{\ell}
        $$
        \item Multiplication of these SO(3) representations corresponds to multiplication of their spherical functions (dropping high-frequency terms where needed)
    \end{itemize}
    \item In addition to taking non-scalar values, the weights are also spatially varying \emph{functions}, depending on the relative vector between the communicating nodes. The spatial variation of these weights is also represented using a spherical harmonic decomposition, with radial dependence.
    \begin{itemize}
        \item Therefore, the weight functions (and thus the model) are \emph{equivariant} if rotating the evaluation vector in 3D space corresponds to ``rotating'' the weight value.
    \end{itemize}
\end{itemize}

To compute the product of spherical functions $f$ and $g$ with harmonic decompositions $f_{\ell_1,m_1}$ and $g_{\ell_2,m_2}$, Equiformer uses tensor products based on Clebsch–Gordan coefficients\cite{cg_coeff1960} $C_{\ell_1,\ell_2}^{\ell_3}$, which combine the components in the correct way:
\begin{align}
[f_{\ell_1} \otimes g_{\ell_2}]_{\ell_3,m_3} = \sum_{m_1, m_2} C_{\ell_1,m_1;\,\ell_2,m_2}^{\ell_3,m_3}\,f_{\ell_1,m_1}\,g_{\ell_2,m_2}.
\end{align}

Of course, what distinguishes Equiformer from an ordinary equivariant message passing network is that Equiformer uses the above operations to build an equivariant attention mechanism, so that each node and each head can pay different amounts of attention to different neighbor nodes.

EquiformerV2~\cite{liao2023equiformerv2} enhances the original Equiformer architecture. It replaces the SO(3)-equivariant convolutions with eSCN convolutions, reducing computational complexity from \(O(L_\text{max}^6)\) to \(O(L_\text{max}^3)\), enabling scaling to higher-degree ($L=6$) representations~\cite{passaro2023escn, liao2023equiformerv2}.

EquiformerV2 achieved state-of-the-art results on large-scale datasets (e.g., OC20/OC22), which use force and energy of individual molecules as the training target. However, EquiformerV2 is not equipped to predict solubility.

\subsection{Contribution: Solvaformer}
\label{sec:contribution}
In our work, we modify the EquiformerV2 architecture to enable solubility prediction. Instead of receiving one molecule as input, Solvaformer receives multiple molecules (a solute and one or more solvents). EquiformerV2 only uses equivariant attention, whereas in Solvaformer there are two types of attention: equivariant attention between intramolecular atoms and scalar attention between intermolecular atoms.

We don't know the molecules' spatial relationship to each other (in fact, this is often not well specified for solutions), therefore the symmetries of the model include one independent copy of $SE(3)$ for each molecule. I.e., it should be possible to move one of the molecules without affecting the output; therefore, 3D vectors between the atoms of one molecule and the atoms of another are not meaningful, and should not be involved in any computations. This is why we use only scalar attention between molecules: keys and queries are computed using \emph{only} the scalar part of node features.

Equivariant and scalar attention modules aggregate messages from within-molecule and from other-molecules. Suppose $i$ indexes a destination atom, $j$ indexes another atom in the same molecule, and $\zeta$ indexes an atom in the other molecule, with embeddings $x_i$, $x_j$, and $x_\zeta$. To compute the message incident on atom $i$, we compute the messages from equivariant attention and scalar cross-attention, and then sum them. For the equivariant attention, Equiformer computes message tensor values $v^{e}_{ij}$ using tensor products between $x_i$ and $x_j$, and similarly projects from $x_i \otimes x_j$ down to scalar features $f^{(0)}_{ij}$, which it then passes through a layer norm and activation to produce a logit
\begin{align}
z^e_{ij} = \textrm{LeakyRELU}(\textrm{LayerNorm}(f^{(0)}_{ij})). 
\end{align}
(Note that this is more general than dot product attention.) The scalar cross-attention is a simple implementation of traditional dot product attention, with the message values $v^s_\zeta$, and the key and query vectors, $k_\zeta$ and $q_i$ computed from the scalar part of the embedding, $x^{(0)}_\zeta$, by linear maps. Then the logits are $z^s_{i\zeta} = \langle{q_i, k_\zeta}\rangle$. The messages are aggregated the same way for both, with the caveat that equivariant messages are tensorial, while scalar messages are purely scalar quantities:
\begin{gather}
m^{e}_i = \sum_j \mathrm{softmax}_j (z^e_{ij})\;\, v^{e}_{ij} \qquad
m^s_i = \sum_\zeta \mathrm{softmax}_j (z^s_{i\zeta})\;\, v^s_{i\zeta} \\
\mu_i = m^{e}_i + m^s_i
\end{gather}
where, of course, the sum happens only in the scalar degree.

Solvaformer is trained using a multi-task objective (solubility and solvation energy) which enables us to utilize large data generated by COSMO-RS calculations (CombiSolv-QM) and experimental solubility measurements (BigSolDB 2.0). Solvaformer achieves state-of-the-art performance of solubility prediction on our benchmark dataset. Additionally, the attention weights of the model can be used to generate explainability charts that indicate which molecular relationships are driving predictions.

\begin{figure}
    \centering
    \includegraphics[width=\linewidth]{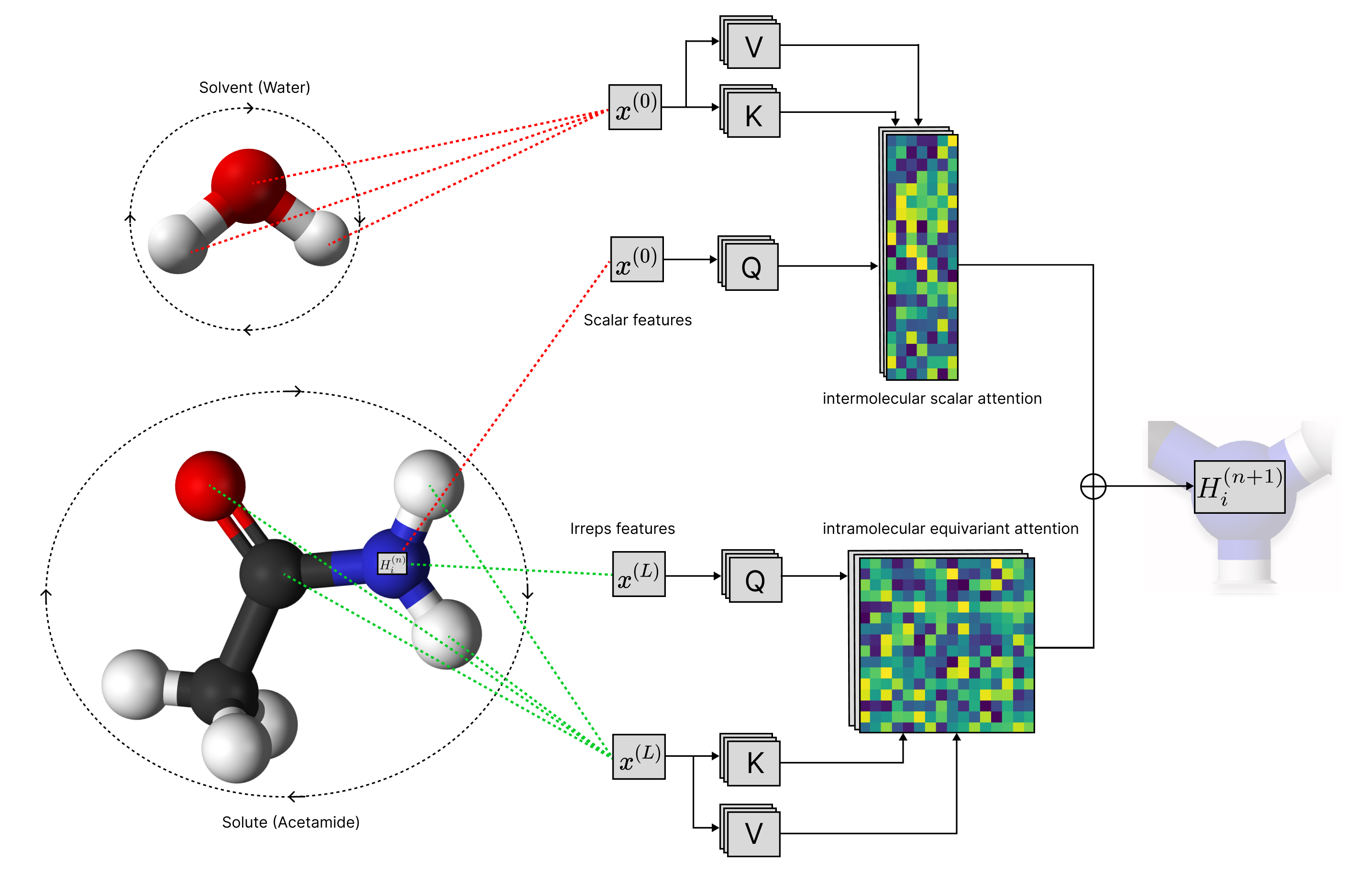}
    \caption{Example of Solvaformer performing an update of the hidden representation of a nitrogen atom ($H_i$) in a single layer for inputs water and acetamide. Irreps features contain relative 3D positions defined by spherical harmonics and are collected only from intramolecular atoms.  Since only scalar features are received from the atoms of the solvent, the hidden representation is constructed assuming that intermolecular atoms are not SE(3) equivariant.}
    \label{fig:diagram}
\end{figure}

\section{Methods}
\label{sec: methods}
\subsection{Data}
\subsubsection{BigSolDB 2.0}
We utilized the BigSolDB 2.0 dataset, a comprehensive solubility resource comprising 103{,}944 experimentally measured solubility values for 1{,}448 unique organic solutes in 213 solvents, across a temperature range of 243--425~K. These values were manually curated from 1{,}595 peer-reviewed publications and standardized into a machine-readable format including SMILES representations for both solutes and solvents. LogS values (log molar solubility in mol/L) were calculated using solvent densities either from experimental measurements or interpolated via linear models where necessary. The dataset spans aqueous and non-aqueous solvents, including common organic media such as ethanol, acetone, and ethyl acetate, enabling broad coverage for solubility prediction tasks \cite{Krasnov2023, krasnov2025bigsoldb}.

To ensure the quality of the data used for model development, we applied the following filtering criteria:
\begin{itemize}
    \item Canonicalized the SMILES of both solutes and solvents using RDKit.
    \item Removed entries containing bimolecular solutes or multi-component species.
    \item Excluded all metal-containing and ionic solute entries.
    \item Discarded entries lacking a LogS value.
    \item Discarded all duplicate entries.
\end{itemize}
The dataset had 6591 duplicated entries. LogS between duplicated entries had an average standard deviation of $0.0974$, which represents an intrinsic limit to the performance of our models (figure \ref{fig:error_in_data}).
Following data filtering, we split the dataset into training and test sets using chemical space-aware clustering:
\begin{itemize}
    \item Solutes were clustered using the Butina algorithm based on Tanimoto similarity of Morgan fingerprints (radius = 2).
    \item From these clusters, we sampled solutes across the chemical space to form a structurally diverse test set (10\% of data).
\end{itemize}

The final split consisted of 82{,}758 solute-solvent measurements for training and 9{,}250 for testing, with 1{,}142 unique solutes in the training set and 126 in the test set (See figures \ref{fig:train-test-split}, \ref{fig:train-test-logs}). 

This stratified, diversity-aware split enables robust benchmarking of model generalization to solute structures.

\subsubsection{CombiSolv}

The CombiSolv-QM dataset~\cite{vermeire2020transfer} provides quantum-mechanically computed solvation free energies for approximately one million solvent–solute pairs. These values were derived using COSMO-RS theory via the COSMOtherm software, covering 11{,}029 solutes and 284 solvents. All solvation energies (\(\Delta G_\text{solv}\)) were calculated at 298~K using a conformer-aware protocol that includes DFT-based geometry optimization followed by chemical potential analysis in solution.

In our work, CombiSolv-QM was used in conjunction with the BigSolDB 2.0 dataset to train the Solvaformer model. To enable learning from both experimental and computational data, we implemented an alternating batch training scheme: each mini-batch was sampled from either CombiSolv-QM or BigSolDB 2.0. The model was trained with two separate prediction tasks: one for experimental solubility values (LogS) and one for calculated solvation free energies (\(\Delta G_\text{solv}\)). The training set of CombiSolv-QM was filtered for solutes in the BigSolDB 2.0 test set.

This dual-target, alternating-batch strategy allowed the model to generalize across both experimental and theoretical chemical space while maintaining fidelity to each target type. It also served as an effective form of multi-task learning, allowing the model to benefit from the scale and consistency of CombiSolv-QM and the real-world relevance of BigSolDB 2.0.

\subsection{Models}
\subsubsection{XGboost models}
\label{sec: xgboost}

We use XGBoost to predict solubility using a variety of embedding methods. For \textbf{DFT} features, we first generated conformers using RDKit ETKDGv3 through WEASEL~1.12. Conformers were then optimized using GFN2-xTB, and the five most stable conformers (or those covering 90\% of the Boltzmann population at the xTB level) were subsequently subjected to DFT calculations at the wB97X-V/def2-TZVP level of theory in the gas phase. The energies derived from these DFT calculations were used to apply Boltzmann weights to the resulting molecular features. Features were calculated for both the solute and the solvent molecules. The most stable conformer was further evaluated with a COSMO-RS calculation using ORCA~6.0\cite{neese2012orca} (license: EULA) to obtain the free energy of solvation.

We also generated \textbf{Circular Fingerprints (CFP; Morgan fingerprints)}\cite{morgan1965generation} with RDKit (license: CC BY-SA 4.0), using a radius of 2 and a fingerprint size of 2048 bits, with count simulation disabled, chirality excluded, bond types enabled, ring-membership information included, default count bounds, and without restricting to nonzero invariants.

In addition, we produced learned embeddings using \textbf{MegaMolBART (MMB)}, a large language model trained on 1.5~billion SMILES from the ZINC-15 dataset~\cite{irwin2022chemformer, NVIDIA_MegaMolBART_2024} (license: Apache-2.0) and previously shown to be effective for molecular property prediction~\cite{moayedpour2024representations}. From the pretrained model we extracted 512-dimensional embeddings and used them directly as inputs to XGBoost.

Furthermore, the \textbf{SolvBERT} model was initially trained for temperature independant solubility. To adapt it for our purposes we first trained the model on the CombiSolv and BigSolDB 2.0 training data (Training followed the procedure described here: https://github.com/su-group/SolvBERT \cite{yu2023solvbert}; license: CC-BY 4.0) and then took the embeddings from the pre-trained model and trained an XGBoost regressor with temperature as an additional feature to the model.

Using these feature sets, we trained a standard XGBoost regressor~\cite{chen2016xgboost} (license: Apache-2.0) with \texttt{n\_estimators}=500, \texttt{learning\_rate}=0.1, and \texttt{max\_depth}=6.

\subsubsection{Message passing neural network (MPNN)}
We adopted a MPNN architecture variation (gated graph neural network; GG-NN+set2set) specifically developed for molecular property prediction~\cite{gilmer2017mpnn} (license: Apache-2.0) by modifying the GG-NN message passing architecture. The model consists of three stages, namely the message-passing phase followed by an interaction phase, and finally the prediction phase. In the message-passing phase, features of the immediate neighboring nodes are aggregated into a node's contextual information via unidirectional edge networks. This process is repeated for \textit{n} message passing steps, which is a hyperparameter tuned when training our model. In the interaction phase, an interaction map is built by performing a matrix multiplication on the aggregated solute and solvent feature tensors. Solute-solvent interactions are then resolved by mapping solute features on to the solvent tensor and vice versa. Finally, the updated solute and solvent tensors are concatenated along with 'temperature' as an external input to create the final features' tensor. The final features' tensor is passed through three ReLU activation layers before correlating with the target, logS in this case. Dimensions of the activation layers are hyperparameters that are tuned when training the model.

\subsubsection{Solvaformer}
\label{sec:solvaformer}

We trained \textit{Solvaformer}, an equivariant graph transformer model, to predict solubility using a combined dataset of BigSolDBv2 and CombiSolv-QM, sampled in equal ratio via alternating batches. The model was trained with a batch size of 6 across 8 devices for up to 1000 epochs, with early stopping based on a patience of 20 and a minimum delta of 0.001. Solvaformer consists of 8 layers with 10 attention heads, 128-dimensional spherical channels, and 96-dimensional hidden dimensions in both attention and feedforward networks. It uses SE(3)-equivariant operations with angular momentum up to \( l = 6 \), and includes solvent-solvent attention and edge features. Regularization includes alpha dropout (0.13), drop path (0.07), and projection dropout (0.06). The model predicts both solubility and solvation energy using separate outputs and is optimized with mean squared error loss and a learning rate of \(3 \times 10^{-6}\). All hyperparameters were selected based on a hyperparameter optimization experiment, the details of which are provided in the Appendix (figure \ref{fig:hparam}). Training progress was tracked using Weights \& Biases.

\subsubsection{EquiformerV2}
We do an ablation test for our Solvaformer model from the original EquiformerV2. Equiformer V2 was trained and evaluated with the identical hyper-parameter configuration defined for Solvaformer (Section \ref{sec:solvaformer}). The model was adapted to accept two molecular graphs and to optimise simultaneously for solvation free energy and LogS. The only architectural change relative to Solvaformer is the removal of the cross-molecular scalar-attention mechanism; intramolecular message passing is retained, but explicit intermolecular communication is disabled.

\subsection{Compute Resources}
\label{sec: compute}
\subsubsection{Training}
The experiments presented in this work require varying levels of computational resources. While most models can be trained on CPU-only machines, training MPNN, Solvaformer, and EquiformerV2 necessitates GPU acceleration. These models were trained on an Nvidia DGX cloud cluster with 24 H100 GPUs.

The DFT features were generated on a machine with 64 CPUs and 256GB of memory in 36 hours. Everything was run in parallel using a single core for each calculation, the smaller molecules take a couple minutes and the larger can take upward of 10 hours.
MPNN was trained on a single 80GB Nvidia H100 GPU, utilizing only 10\% of the available GPU memory and 16 CPU cores. It converged in 126 minutes. In contrast, Solvaformer required significantly more compute: it utilized approximately 80\% of the same GPU’s memory, 32 CPU cores, and took around 160 hours to converge. EquiformerV2 had similar resource demands to Solvaformer.
\subsubsection{Inference}
In contrast to training, inference times for the evaluated models vary considerably. The XGBoost-DFT model requires as much time for inference as it does for training (per molecule). Predictions for very large molecules can potentially take up to 10 hours. This makes it impractical for real-world applications, and its results are instead presented as a benchmark of a less deployable but gold standard predictive approach. The remaining models achieve markedly faster inference, with computations completing within minutes. For example, Solvaformer can determine the solubility of 9,250 test samples in under two minutes when running on an Nvidia H100 GPU, and comparable throughput was observed for the other deep learning architectures.

\section{Results}
\begin{table}
\label{tab:model_performance}
\centering
\caption{Model performance metrics on BigSolDBv2 test set}
\begin{tabular}{lrrrrrr}
\toprule
 & MAE & MSE & RMSE & $R^2$ & Pearson & Spearman \\
Model &  &  &  &  &  &  \\
\midrule
\textit{XGBoost-DFT }& 0.621 & 0.650 & 0.806 & 0.499 & 0.722 & 0.722 \\
\hline \\
XGBoost-CFP & 0.749 & 0.889 & 0.943 & 0.315 & 0.592 & 0.561 \\
XGBoost-MMB & 0.710 & 0.831 & 0.912 & 0.360 & 0.616 & 0.591 \\
SolvBERT & 0.871 & 1.231 & 1.110 & 0.052 & 0.247 & 0.222 \\
MPNN & 0.668 & 0.746 & 0.864 & 0.425 & 0.724 & 0.693 \\
EquiformerV2 & 0.691 & 0.858 & 0.926 & 0.345 & 0.626 & 0.606 \\
Solvaformer & 0.643 & 0.700 & 0.837 & 0.460 & 0.694 & 0.677 \\
MPNN+Charges & 0.629 & 0.667 & 0.817 & 0.486 & 0.721 & 0.710 \\
\bottomrule
\end{tabular}
\end{table}

Table~\ref{tab:model_performance} summarizes the predictive performance of six models on the BigSolDBv2 test set. XGBoost-DFT is also included here for comparison, however it is set as apart as we note that it's very high inference time makes it an impractical solution. \textbf{Solvaformer} demonstrates strong overall performance, with an MAE of 0.643, an RMSE of 0.837, and an $R^2$ of 0.460. These values are close to those of the XGBoost-DFT baseline, which attains the lowest MAE (0.621), the lowest RMSE (0.806), and the highest $R^2$ (0.499) among all evaluated methods.

The \textbf{MPNN} baseline is also competitive, with an MAE of 0.668 and an RMSE of 0.864, and it achieves the highest Pearson correlation coefficient (0.724) on this split. In terms of rank correlations, MPNN reaches a Spearman value of 0.693, which is comparable to Solvaformer’s 0.677.

Other approaches, including XGBoost-CFP and XGBoost-MMB, yield higher errors and lower correlations than the top-performing methods, and \textbf{EquiformerV2} similarly lags behind across the reported metrics. \textbf{SolvBERT} exhibits the weakest agreement with ground truth, with an $R^2$ of 0.052 and a Pearson correlation of 0.247.

Overall, the results indicate that Solvaformer provides the strongest performance among the learned models, with MPNN close behind. Solvaformer most closely approaches the performance of the DFT assisted regression.

\subsection{Explainability Case Study: Distinguishing Intra- vs. Intermolecular Hydrogen Bonds}

A key advantage of Solvaformer is its ability to interpret the complex 3D relationships that govern molecular interactions. To provide a compelling demonstration, we analyzed the model's attention maps for the solubility of two isomers in water: salicylic acid (\textit{ortho}-hydroxybenzoic acid) and 4-hydroxybenzoic acid (\textit{para}-hydroxybenzoic acid). This pair represents a classic chemical challenge where the change in substituent position dictates whether hydrogen bonding is internal (intramolecular) or external (intermolecular).

\begin{figure}[h]
    \centering
    \begin{tabular}{cc}
        \includegraphics[width=0.45\linewidth]{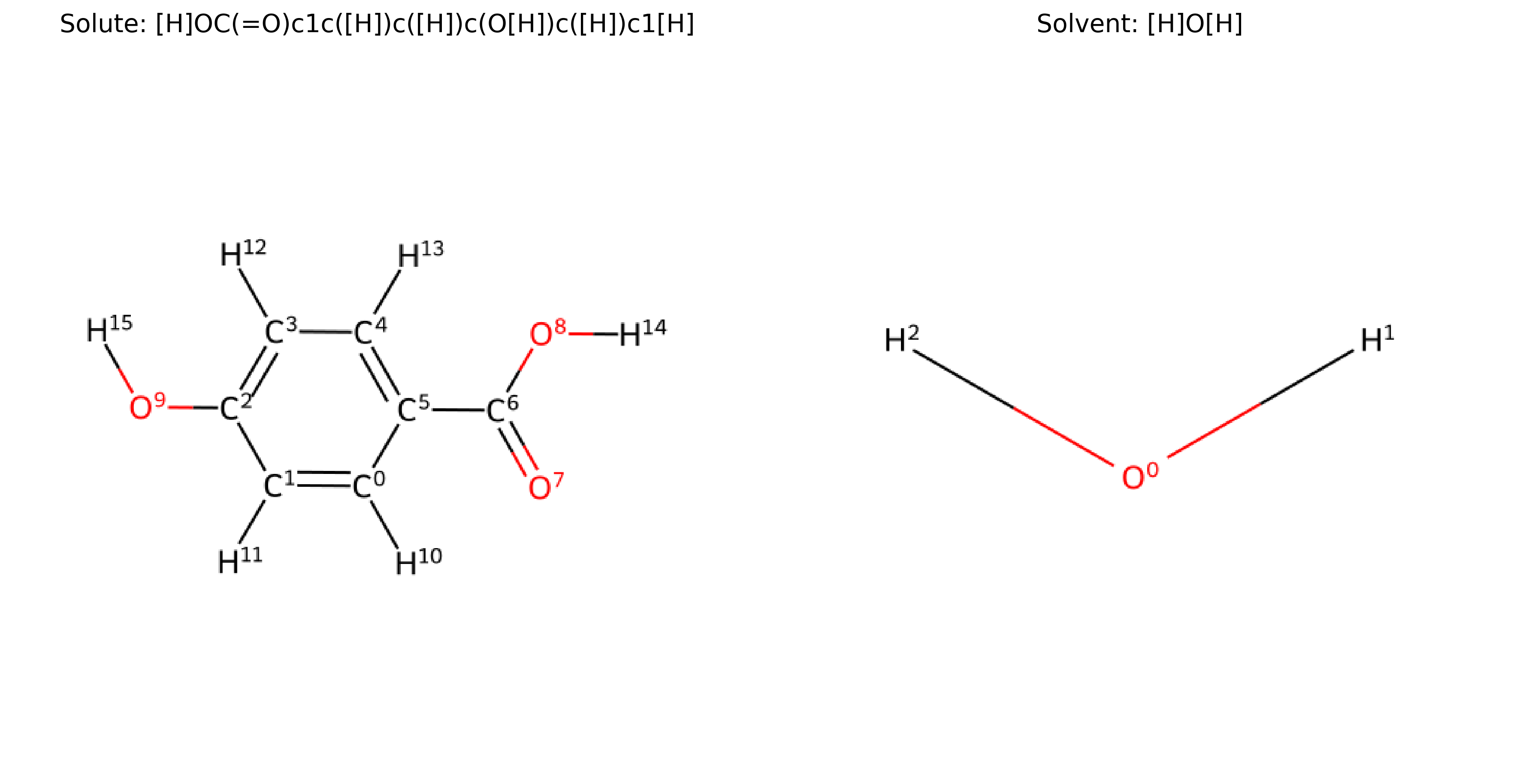} &
        \includegraphics[width=0.45\linewidth]{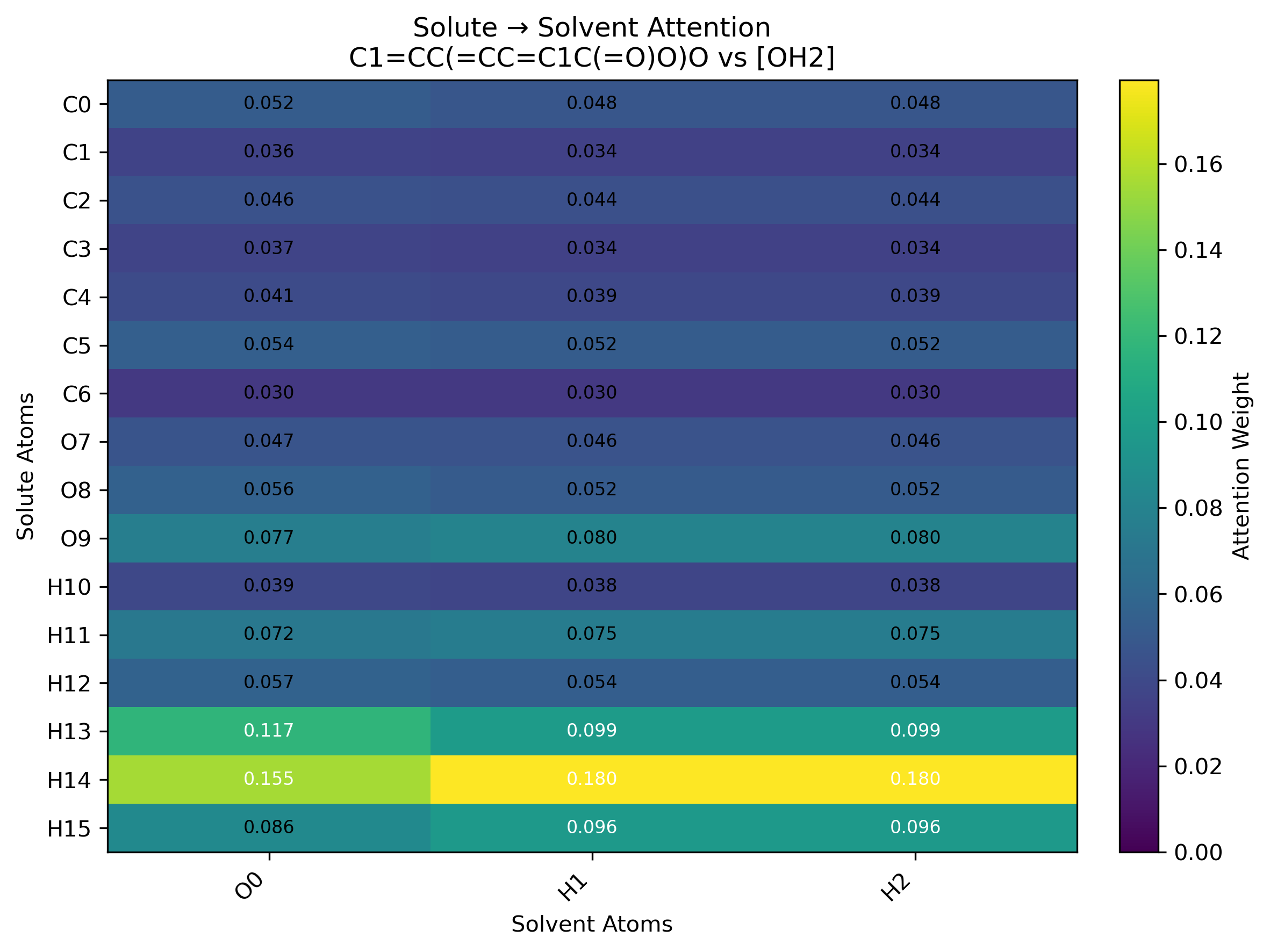} \\
        (a) 4-hydroxybenzoic acid (\textit{para}) & (b) H15 shows intermolecular attention \\
        \\
        \includegraphics[width=0.45\linewidth]{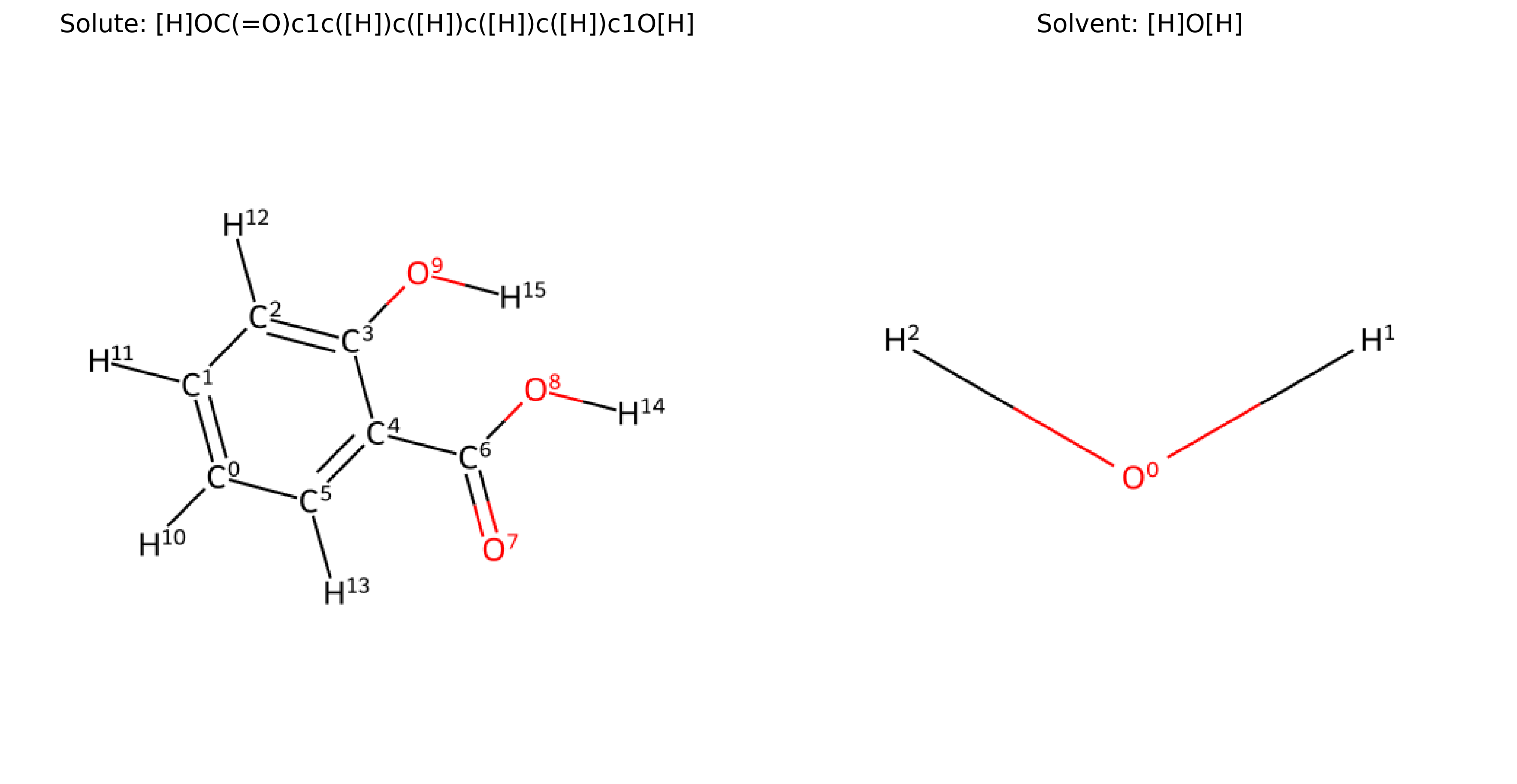} &
        \includegraphics[width=0.45\linewidth]{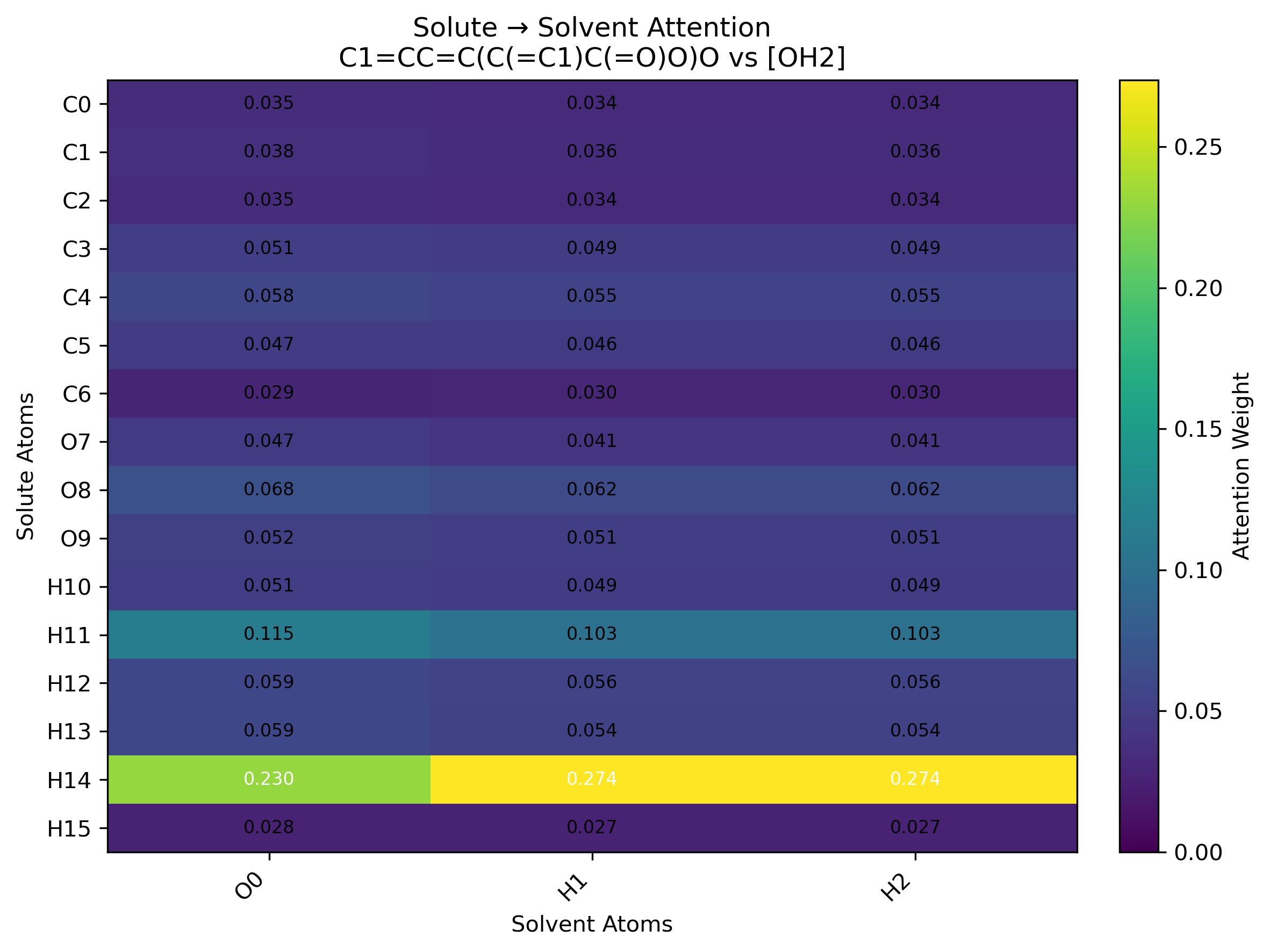} \\
        (c) Salicylic acid (\textit{ortho}) & (d) Intermolecular attention from H15 is absent \\
    \end{tabular}
    \caption{Solute-to-solvent attention maps demonstrating Solvaformer's chemical intuition. (b) For the \textit{para} isomer, the hydroxyl proton (H15) shows clear attention to water, indicating intermolecular H-bonding. (d) For the \textit{ortho} isomer, this attention from H15 disappears, correctly implying it is occupied in a dominant intramolecular H-bond.}
    \label{fig:isomer_attention}
\end{figure}

The key atom for this comparison is the hydroxyl proton, labeled H15 in our structures. In 4-hydroxybenzoic acid, H15 is exposed and free to form intermolecular hydrogen bonds with water, enhancing solubility. In salicylic acid, however, the adjacent geometry allows H15 to form a strong intramolecular hydrogen bond with the carbonyl oxygen. This internal bond makes H15 unavailable for solvent interactions, thus lowering solubility. The attention maps in Figure~\ref{fig:isomer_attention} show that Solvaformer captures this distinction.

For the \textit{para} isomer (Figure~\ref{fig:isomer_attention}b), the hydroxyl proton H15 is geometrically unhindered. The attention map correctly reflects this by showing a distinct interaction between H15 and the atoms of the water molecule. This signal is direct evidence that the model identifies H15 as an active site for intermolecular hydrogen bonding.

In stark contrast, the attention map for salicylic acid (\textit{ortho}) (Figure~\ref{fig:isomer_attention}d) shows that the attention between the hydroxyl proton H15 and the solvent is effectively zero. This absence of interaction is the critical finding. It demonstrates that the model has learned that H15 is "occupied" by the intramolecular hydrogen bond with the nearby carbonyl oxygen and is therefore unavailable to bond with water.

This case study proves that Solvaformer is not merely correlating features but is learning physically meaningful, 3D-aware principles of solvation chemistry. The ability to distinguish between competing bonding scenarios based on geometry is a sophisticated piece of chemical reasoning that makes the model's predictions both more accurate and highly interpretable.

\section{Discussion}
\paragraph{Main findings.}
Table~\ref{tab:model_performance} summarizes results on BigSolDBv2. \textbf{Solvaformer} delivers competitive---and in several metrics, best---performance among all methods while preserving strong geometric inductive bias and interpretability. Relative to the geometry-only baseline (\textit{EquiformerV2} without cross-molecular scalar attention), Solvaformer consistently improves error (MAE/RMSE) and correlation (Pearson/Spearman), indicating that coupling \emph{intramolecular} SE(3)-equivariant attention with \emph{intermolecular} scalar attention better captures solute--solvent interactions. 

\paragraph{Comparison to baselines.}
DFT-assisted gradient boosting (XGBoost-DFT) is a strong classical baseline, but it is not a practical solution for this problem as it has a high runtime at inference (up to 10 hours). Fingerprint-based models (XGBoost-CFP) and language-model embeddings (XGBoost-MMB) trail the top methods, suggesting that fixed 2D encodings or generic sequence embeddings underutilize solution-phase geometry. The MPNN baseline remains robust but lacks explicit symmetry handling and offers less transparent attributions. SolvBERT underperforms on this split, aligning with reports that concatenated-SMILES encoders can struggle with out-of-domain solute--solvent combinations.


\paragraph{Limitations.}
\label{sec: limitations}
While the model demonstrates different attention to geometric isomers (e.g., cis/trans or $E/Z$), the solubility prediction is relatively insensitive to these changes, so differences that arise primarily from configuration may be underrepresented. Our evaluation is based on a single, chemistry-aware train–test split; because each training run requires multiple days, exhaustive $k$-fold cross-validation was impractical, limiting our ability to characterize variance across splits. The measured solubility labels are drawn from a meta-analysis of external measurements, introducing potential heterogeneity from differing experimental protocols and curation practices. In addition, individual measurements carry inherent experimental error (figure \ref{fig:error_in_data}), which propagates into both the training targets and the reported evaluation metrics.

\paragraph{Future directions.}
As mentioned in the section above, we observe that solubility predictions between geometric isomers could be improved. We see that this can be a critical factor in solubility optimization~\cite{schoen2014separation}. The current training data includes few such isomeric pairs with distinct measured solubilities. To further enhance model performance, we plan to curate an internal dataset enriched with these cases. A controlled test dataset will also give us a defined measurement of uncertainty. This would provide a more challenging and reliable benchmark to better evaluate geometric sensitivity.
Additionally, our current data preprocessing pipeline generates only a single 3D conformer per solubility measurement. In future work, we aim to enhance this step by generating multiple conformers per molecule to augment the dataset and better capture conformational variability.

\paragraph{Takeaway.}
Our study suggests that Solvaformer offers a practical balance of accuracy, scalability, and interpretability for solution-phase modeling. By combining geometry-aware design with a mixed data training regimen and providing mechanistically plausible attributions, the approach supports both high-throughput screening and hypothesis-driven analysis. With MPNN as a strong comparator and XGBoost-DFT as an upper bound under heavier computation, this toolkit is well suited for reliable solubility prediction in applied settings.
\newpage

\printbibliography

\newpage
\appendix
\section{Additional Tables and Figures}

\begin{figure}[!htb]
    \centering
    \includegraphics[width=0.9\linewidth]{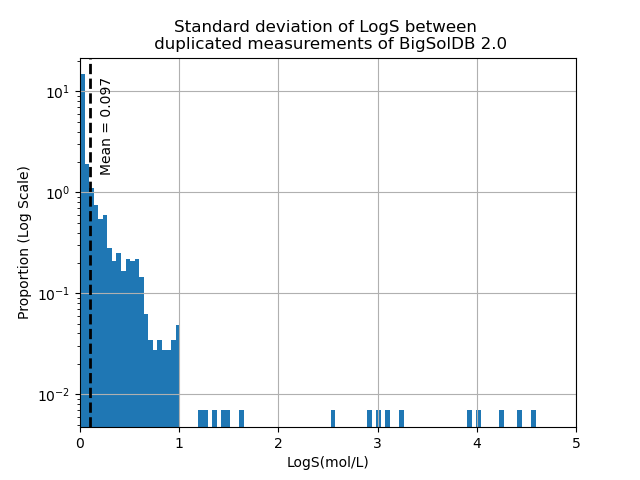}
    \caption{6591 measurements in BigSolDB 2.0 had duplicated measurements from seperate sources (same solute, solvent, temperature but measured in a different labratory and have different measured solubility). We removed these measurements from the dataset. Here we measure the standard deviation within groups of duplicated measurements and plot the distribution. This provides an estimate of the precision of experimental measurements for solubility and hence lower bound for error rate prediction of our dataset.}
    \label{fig:error_in_data}
\end{figure}

\begin{table}[!htb]
    \centering
    \caption{DFT calculated molecular features}
    \begin{tabular}{cl}\hline
         \textbf{Feature name}& \textbf{Feature Description}\\\hline
         Dipole\_Moment\_Debye& Dipole moment in Debye\\\hline
         LUMO\_E\_Eh& Energy of Lowest unoccupied molecular orbital (LUMO) in Hartrees\\\hline
 LUMOX\_E\_Eh&Energy of LUMO - X in Hartrees\\\hline
 HOMO\_E\_Eh&Energy of Lowest occupied molecular orbital (HOMO) in Hartrees\\\hline
 HOMOX\_E\_Eh&Energy of HOMO - X in Hartrees\\\hline
 HOMO\_LUMO\_gap&Energy difference between HOMO and LUMO\\\hline
 Dispersion\_correction&Dispersion correction calculated with VV10 nonlocal van der Waals correlation\\\hline
 Cavity\_Volume&CPCM cavity volume in cubic angstroms\\\hline
 Cavity\_Surface\_area&CPCM solvent-accessible surface in squared angstroms\\\hline
 Surface\_Charge\_CPCM&Total apparent surface charge distribution calcualted by CPCM\\\hline
 C\_charge\_total&Sum of all Hirshfeld charge on all carbon atoms\\\hline
 O\_charge\_total&Sum of all Hirshfeld charge on all oxygen atoms\\\hline
 N\_charge\_total&Sum of all Hirshfeld charge on all nitrogen atoms\\\hline
 H\_charge\_total&Sum of all Hirshfeld charge on all hydrogen atoms\\\hline
 Het\_charge\_total&Sum of all Hirshfeld charge on all heteroatoms\\\hline
 energy\_kcal\_mol&Electronic energy of the system in kcal/mol\\\hline
 dGs&Free energy of solvation as calculated by open COSMO-RS through ORCA6\\ \hline
    \end{tabular}
\end{table}
\begin{figure}[!htb]
\includegraphics[width=\linewidth]{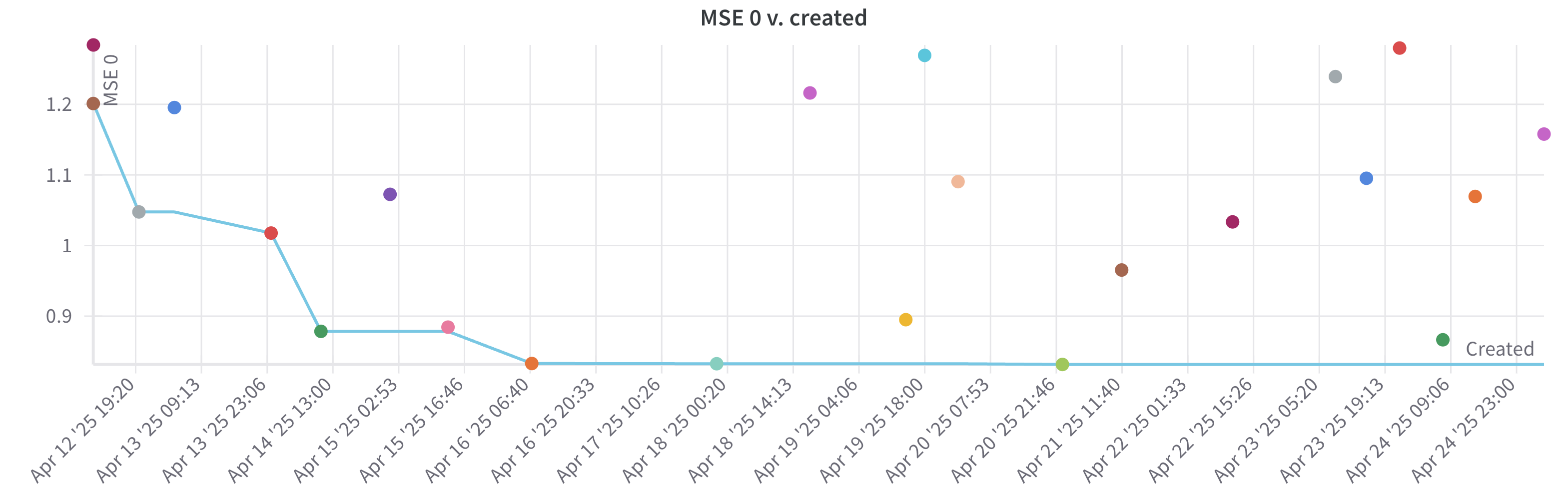}
\includegraphics[width=\linewidth]{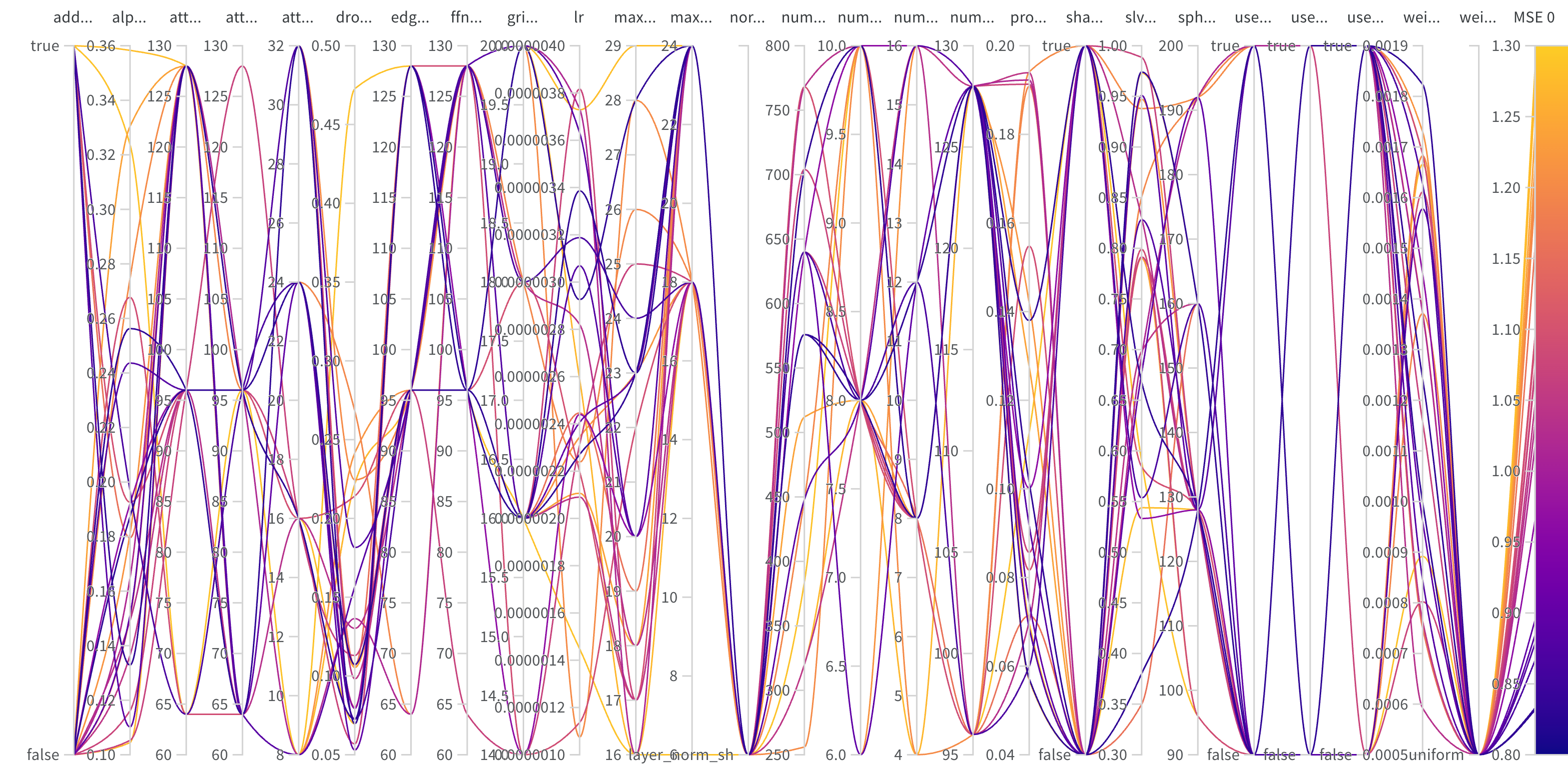}
\label{fig:hparam}
\caption{Hyperparameter tuning of Solvaformer. We ran a total of 23 different runs. WandB agents selected hyperparameters of successive runs using Bayesian optimization where performance was measured by MSE on the BigSolDB2.0 validation set.}
\end{figure}

\begin{figure}[!htb]
    \centering
    \includegraphics[width=0.5\linewidth]{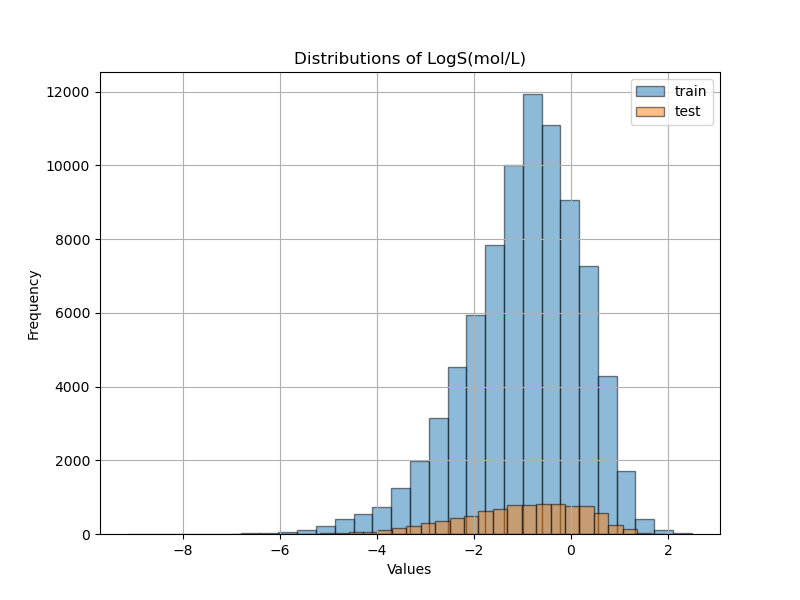}
    \caption{Distribution of measured logS in the train-test split.}
    \label{fig:train-test-logs}
\end{figure}
\begin{figure}[!htb]
    \centering
    \includegraphics[width=0.5\linewidth]{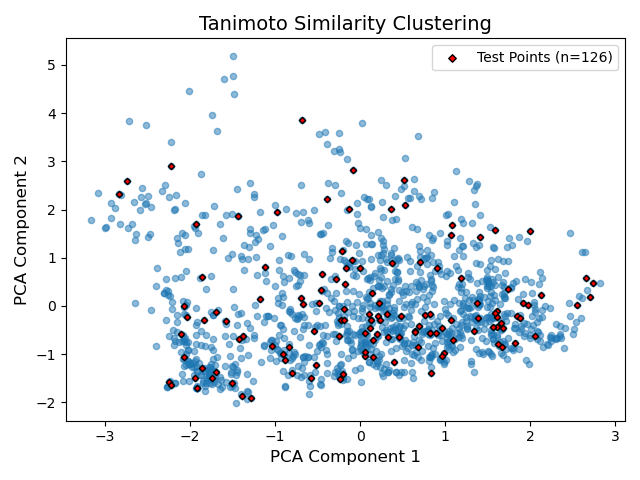}
    \caption{Butina clustering of tanimoto similarity of all unique solutes in BigSolDBv2}
    \label{fig:train-test-split}
\end{figure}

\newpage
\section{Data Availability}
\label{sec: data}
All the raw data used to train and test the models is publicly available and can be found here:
\begin{itemize}
    \item BigSolDB2.0~\cite{krasnov2025bigsoldb}
        \subitem link: \href{https://zenodo.org/records/15094979}{https://zenodo.org/records/15094979}
        \subitem version: Published March 27, 2025 | Version v1
        \subitem license: CC-BY 4.0
    \item CombiSolv-QM~\cite{vermeire2020transfer}:            
        \subitem link: \href{https://zenodo.org/records/5970538}{https://zenodo.org/records/15094979}
        \subitem version: Published July 1, 2022 | Version v1.2
        \subitem license: CC-BY 4.0
\end{itemize}

\end{document}